\makeatletter
\declare@file@substitution{revtex4-1.cls}{revtex4-2.cls}
\makeatother

\documentclass[twocolumn]{aastex631}

\usepackage{xspace}
\usepackage[flushleft]{threeparttable}
\usepackage{color}
\usepackage{url}
\usepackage{hyperref}
\usepackage{graphicx}

\newcommand {\flux}{erg\,s$^{-1}$\,cm$^{-2}$\xspace}
\newcommand {\lum}{erg\,s$^{-1}$\xspace}
\newcommand {\swift}{{Swift}\xspace} 
\newcommand {\ixpe}{{IXPE}\xspace}
\newcommand {\nustar}{{NuSTAR}\xspace}

\shorttitle{X-ray polarimetry of 4U~1820$-$303}
\shortauthors{Di Marco et al.}

\begin{document}

\title{First detection of X-ray polarization from the accreting neutron star 4U~1820$-$303}

\correspondingauthor{Alessandro Di Marco}
\email{alessandro.dimarco@inaf.it}
\author[0000-0003-0331-3259]{Alessandro Di Marco}
\affiliation{INAF Istituto di Astrofisica e Planetologia Spaziali, Via del Fosso del Cavaliere 100, 00133 Roma, Italy}

\author[0000-0001-8916-4156]{Fabio La Monaca}
\affiliation{INAF Istituto di Astrofisica e Planetologia Spaziali, Via del Fosso del Cavaliere 100, 00133 Roma, Italy}
\author[0000-0002-0983-0049]{Juri Poutanen}
\affiliation{Department of Physics and Astronomy,  20014 University of Turku, Finland}
\author[0000-0002-7930-2276]{Thomas D. Russell}
\affiliation{INAF, Istituto di Astrofisica Spaziale e Fisica Cosmica, Via U. La Malfa 153, I-90146 Palermo, Italy}
\author[0000-0002-2701-2998]{Alessio Anitra}
\affiliation{Universit\`a degli Studi di Palermo, Dipartimento di Fisica e Chimica, Emilio Segr\`e, via Archirafi 36 I-90123 Palermo, Italy}
\author[0000-0003-2212-367X]{Ruben Farinelli}
\affiliation{INAF—Osservatorio di Astrofisica e Scienza dello Spazio di Bologna, Via P. Gobetti 101, I-40129 Bologna, Italy}
\author[0000-0003-4216-7936]{Guglielmo Mastroserio}
\affiliation{INAF Osservatorio Astronomico di Cagliari, Via della Scienza 5, 09047 Selargius (CA), Italy}
\author[0000-0003-3331-3794]{Fabio Muleri}
\affiliation{INAF Istituto di Astrofisica e Planetologia Spaziali, Via del Fosso del Cavaliere 100, 00133 Roma, Italy}	
\author[0000-0002-0105-5826]{Fei Xie}
\affiliation{Guangxi Key Laboratory for Relativistic Astrophysics, School of Physical Science and Technology, Guangxi University, Nanning 530004, China}
\affiliation{INAF Istituto di Astrofisica e Planetologia Spaziali, Via del Fosso del Cavaliere 100, 00133 Roma, Italy}
\author[0000-0002-4576-9337]{Matteo Bachetti}
\affiliation{INAF Osservatorio Astronomico di Cagliari, Via della Scienza 5, 09047 Selargius (CA), Italy}
% ...

\collaboration{119}{(IXPE Collaboration)}



\begin{abstract}
This paper reports the first detection of polarization in the X-rays for atoll-source 4U~1820$-$303, obtained with the {Imaging X-ray Polarimetry Explorer} (\ixpe) at 99.999\% confidence level (CL). 
Simultaneous polarimetric measurements were also performed in the radio with the Australia Telescope Compact Array (ATCA).
The \ixpe observations of 4U~1820$-$303 were coordinated with \swift-XRT, NICER, and \nustar aiming to obtain an accurate X-ray spectral model covering a broad energy interval. The source shows a significant polarization above 4\,keV, with a polarization degree of $2.0\%\pm0.5\%$ and a polarization angle of $-55\degr\pm7\degr$ in the 4--7 keV energy range, and  a polarization degree of $10\%\pm 2\%$ and a polarization angle of $-67\degr\pm7\degr$ in the 7--8\,keV energy bin. This polarization also shows a clear energy trend with polarization degree increasing with energy and a hint for a position-angle change of $\simeq 90 \deg$ at 96\% CL around 4 keV. The spectro-polarimetric fit indicates that the accretion disk is polarized orthogonally to the hard spectral component, which is presumably produced in the boundary/spreading layer. We do not detect linear polarization from the radio counterpart, with a $3\sigma$ upper limit of 50\% at 7.25\,GHz.
\end{abstract}


\section{Introduction} \label{sec:intro}

Accreting weakly magnetized neutron stars (NSs) in low-mass X-ray binaries (LMXBs) are among the brightest X-ray sources; they accrete matter via Roche lobe overflow from a stellar companion of mass typically lower than a solar mass \citep{Bahramian2022}. They can be classified according to the shape of their tracks in the X-ray hard-color/soft-color diagram (CCD), or their hard-color/intensity diagram (HID), and their correlated timing in the 1--10\,keV band \citep{vanderklis89,hasinger89}.  The following states are known: (i) the high soft state (HSS) of Z-sources with a luminosity $>10^{37}$\,\lum (near-Eddington X-ray luminosities) exhibiting a wide Z-like shape in the CCD; (ii) the low hard state (LHS) of atoll sources with a luminosity $\simeq$\,10$^{36}$\,\lum showing a single rounded spot in the CCD (island state) typically and having a harder spectrum than HSS; (iii) the HSS of bright atoll sources, having intermediate luminosities ($10^{36}$--$10^{37}$\,\lum), typically following a banana shape figure in the CCD. The banana state has been further divided on the basis of their luminosity into left-lower (lower luminosity), upper (higher luminosity), and lower banana in the middle. 

The spectral and timing properties of these sources provide clues to their emission mechanisms, and their X-ray emission generally is described by two main spectral components: (i) a soft thermal component (blackbody-like emission from the NS surface or from the accretion disk); (ii) a harder component 
associated with the interaction of the accretion disk with the NS surface. 
The interaction region that is coplanar to the accretion disk is called the boundary layer \citep[BL;][]{Shakura88,Popham01}, while the gas layer at the NS surface, extending up to high latitudes, is called the spreading layer (SL) \citep{suleimanov2006,inogamov1999}. 
The properties of the SL/BL region 
(temperature and optical depth) clearly distinguish sources in the LHS from the ones in the HSS, with the former SL/BL much hotter and more transparent with respect to the latter ones \citep{gnarini2022}.
The frequent observation of a reflection component, whose most prominent feature is an iron emission line at $\sim$\,6.5\,keV, suggests Compton reflection by a colder medium (e.g., the outer accretion disk itself) as a further component to take into account, especially in the HSS \citep[see e.g.,][]{Cackett_2008,Titarchuk_2013,Mondal2016,2013A&A...550A...5E}.

The launch of the Imaging X-ray Polarimetry Explorer (\ixpe) in December 2021 \citep{Weisskopf2022,Soffitta_2021}, gave us the opportunity to use X-ray polarization, in addition to spectral and timing information, to disentangle several scenarios. 
The X-ray polarization of weakly magnetized NSs strongly depends on the geometry of the emission region. 
In the LHS sources, it can be produced by a non-spherical slab-like corona \citep[as in the case of BH-LMXB,][]{Haardt93,Poutanen93,PS96,Schnittman2010}, or by the accretion disk \citep{chandrasekhar1960,Loskutov82}.
In the HSS sources, the BL and the SL \citep{Lapidus85} emission can be polarized; also, Comptonization of any seed soft photons in a corona and reflection of the SL radiation off the accretion disk are potential sources of polarized emission  \citep{Lapidus85,suleimanov2006,2010A&A...516A..36D}.

In order to investigate the geometry of the accretion flow in  LMXBs, \ixpe has so far targeted several sources: the Z-source \mbox{Cyg X-2} \citep{Farinelli23}, the Z-atoll transient source XTE J1701$-$462 \citep{Jayasurya2023, Cocchi2023}, and 
two atoll sources, \mbox{GS~1826$-$238} \citep{Capitanio23} and \mbox{GX~9+9} \citep{Chatterjee2023, Ursini2023}. These observations found a higher polarization in the Z-sources than in the atoll ones; moreover, the polarization angle (PA) of Cyg X-2, for the hardest part of the energy spectrum, appears to be aligned with the radio jet. Also, a marginal detection of polarization available for Sco X-1 from OSO-8 \citep{Long1979} and PolarLight \citep{Long2022} appears to show such an alignment between the radio jet and the PA.
An attempt to measure the X-ray polarization of 4U~1820$-$303 was performed by OSO-8 \citep{1984ApJ...280..255H}, but only 99\% confidence level (CL) upper limits of 4.7\% and 10.8\% were obtained at 2.6\,keV and 5.2\,keV, respectively. 
 
\begin{figure}
\centering   
\includegraphics[width=0.9\linewidth]{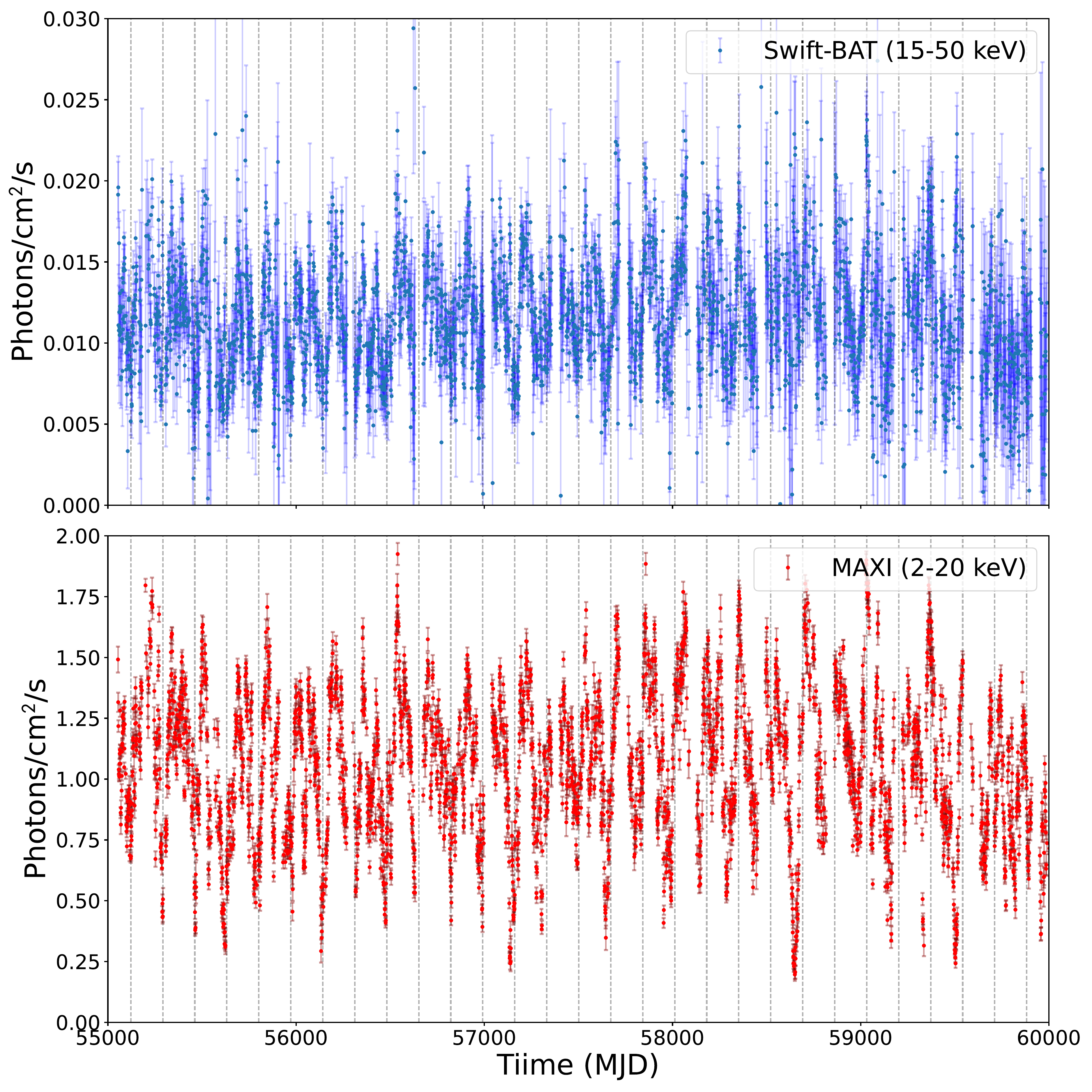}
\caption{Light curves from Swift-BAT monitor in the 15--50\,keV energy band (top) and MAXI in 2--20\,keV energy band (bottom). The dashed  vertical lines identify the 170\,day superorbital period. }
\label{fig:lc_maxi}
\end{figure}

4U~1820$-$303 is an ultra-compact LMXB consisting of a NS, accreting matter via Roche lobe overflow from a He white dwarf. It is located at 0\farcs66 from the center of the globular cluster NGC 6624 \citep{Rappaport87,Shaposhnikov2004}, and its distance was estimated from the GAIA EDR3 to be $D= 8.0 \pm 0.1$ kpc \citep{2021MNRAS.505.5957B}.
4U~1820$-$303 has an orbital period of 685~s \citep{stella87} and a peculiar behavior with intrinsic luminosity variation by a factor of at least 2 along a superorbital $\sim$170\,d period \citep{Zdziarski2007}. \citet{Chou2001} found the flux modulation to be stable in the period 1969--2000 with periodicity 171.0$\pm$0.3\,d, compatible with the RXTE All Sky Monitor (ASM) data \citep{Zdziarski2007}. These variations are also seen in the MAXI and Swift-BAT light curves (see Figure~\ref{fig:lc_maxi}). 
4U~1820$-$303 is the first identified source of type-I X-ray bursts \citep{Grindlay1976}, observed mainly around the flux minima, proving that the observed variability is indeed due to intrinsic accretion rate changes; this is also supported by the strong correlations between the observed flux variations with the source spectral state (in a way typical of atoll-type NS binaries) and the kHz frequency of quasi-periodic oscillations (QPOs) \citep{Smale1997}. 

\begin{figure*} 
\centering
\includegraphics[width=0.8\textwidth]{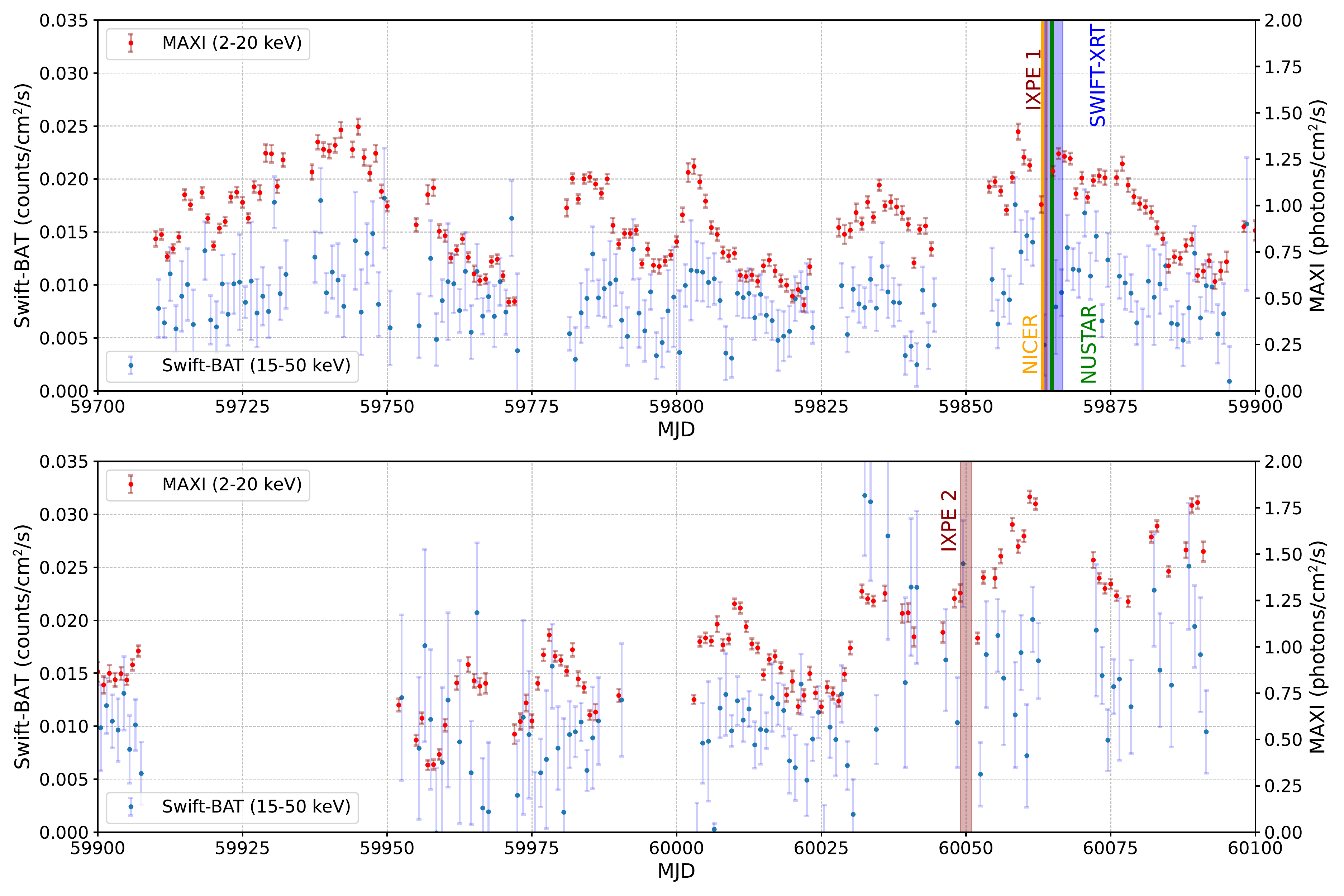}
\caption{Light curves from \swift-BAT in the 15--50\,keV energy band (blue points) and MAXI in the 2--20\,keV energy band (red points), overlaid in dark red are the \ixpe intervals of observation. For the first observation (top panel), where coordinated observations were not strictly simultaneous, vertical bands show the observing period for NICER (orange), \nustar (green), and \swift-XRT (blue), while for the second (bottom panel) they were strictly simultaneous and included in the \ixpe band.}
\label{fig:lc_all}
\end{figure*}

4U~1820$-$303, as typically observed in weakly magnetized NS-LMXBs, shows a spectrum mainly composed of two components: a blackbody or multicolor disk that describes the softer spectral component, and a Comptonization to describe the harder spectral one. Moreover, these sources can show a component due to reflection, which has, as a main feature, the presence of a Gaussian due to the iron K$\alpha$ lines at $\sim$6.5\,keV \citep{Cackett_2008,Titarchuk_2013,Mondal2016}. 4U~1820$-$303 shows a broad iron line \citep{Mondal2016}, which  \citet{Cackett_2008} fitted with a \texttt{diskline} obtaining the disk inclination of $\sim20\degr$.

4U~1820$-$303 is also a known radio emitter \citep[e.g.,][]{2004MNRAS.351..186M,2017A&A...600A...8D,2021MNRAS.508L...6R}. At low X-ray fluxes, the radio spectrum is typically observed to be relatively flat, consistent with a compact radio jet, while, at higher X-ray fluxes \citep{2021MNRAS.508L...6R}, the radio spectrum becomes steep, consistent with either a quenching of the compact jet emission or emission from a transient jet ejecta \citep{2021MNRAS.508L...6R}. Assuming a uniform and ordered magnetic field, linear polarization from a compact (self-absorbed) radio jet is expected to have a maximum degree of $\sim$10\%, while the optically thin ejecta can exhibit values up to $\sim$70\% \citep[see pp. 217--222 in ][]{Longair2011}; however, due to disorder in the magnetic fields, lower values are typically observed \citep[see, e.g.,][]{2014MNRAS.437.3265C}.

\section{X-Ray Observations} \label{sec:handling}

\begin{figure} 
\centering\includegraphics[width=0.8\linewidth]{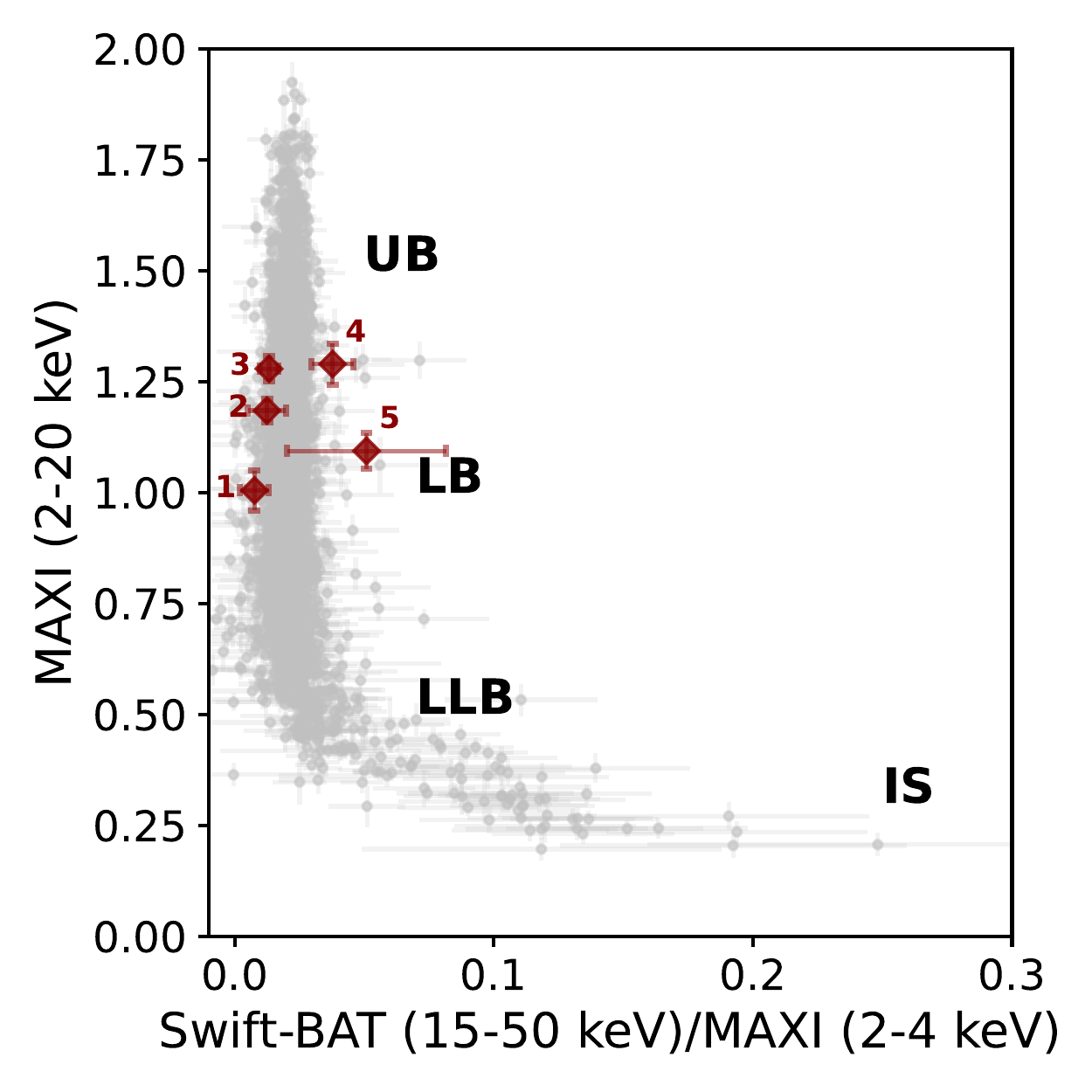}
\caption{Hardness--intensity diagram, where the different states of the source are identified: the island state (IS), where the source has a harder spectrum; the softer banana state with Upper (UB), Lower (LB), and Left Lower (LLB) banana. The red points report the source state during the first and the second \ixpe coordinated observations at the times discussed in the text.}
\label{fig:color}
\end{figure}

\begin{figure*} 
\centering
\includegraphics[width=0.8\textwidth]{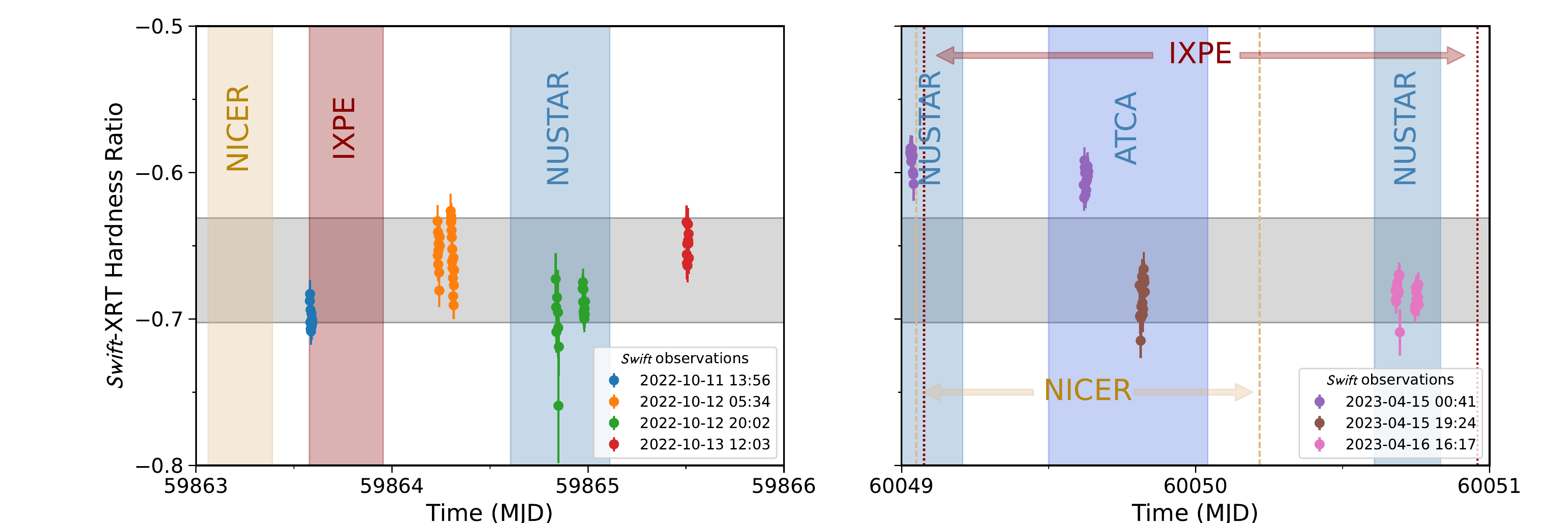}
\caption{\swift-XRT HR in time bins of 100~s during the first (left, we used ObsIDs: 00014980028, 00014980029, 00014980030, and 00014980031) and the second (right, we used ObsIDs: 00014980055, 00014980056, and 00014980057) \ixpe coordinated observations. The horizontal gray band reports the average 99\% CL HR interval during the two observations, the vertical bands and/or horizontal arrows report the time intervals when observations by other observatories were performed.}
\label{fig:hr}
\end{figure*}

\ixpe observed 4U~1820$-$303 in two different periods: on 2022 October 11 from 13:53 to 22:55 UTC for a total exposure of $\simeq$16\,ks per Detector Unit (DU) and from 2023 April 15 01:33 to April 16 23:17 UTC for a total exposure of $\simeq$86\,ks per DU. During these periods, other observatories performed simultaneous observations. To determine the state of 4U~1820$-$303 during the \ixpe observations, we analyzed the daily count rates in the 2--20\,keV energy band from the Monitor of All-sky X-ray Image (MAXI) telescope\footnote{ http://maxi.riken.jp/top/index.html} \citep{MAXI}, and in the 15--50\,keV band from the \swift Burst Alert Telescope (\swift-BAT) Hard X-ray Transient Monitor\footnote{https://swift.gsfc.nasa.gov/results/transients/} \citep{SwiftBAT}. The 4U~1820$-$303 light curves, covering since August 2009 from both observatories, are shown in Figure~\ref{fig:lc_maxi}.  \ixpe observations were performed near the maximum of the flux along the superorbital period, as shown in Figure~\ref{fig:lc_all}.

In Figure~\ref{fig:color}, we show the corresponding hardness-intensity diagram (HID), where the hardness is given by the ratio between \swift-BAT count rate in the 15--50\,keV band and the MAXI count rate in the 2--4\,keV band and the flux is represented by the MAXI count rate in the 2--20\,keV band.
The red points report the state of the source along our coordinated observations: 1 -- on 2022 October 11 by NICER, \ixpe, and \swift-XRT; 2 -- on 2022 October 12  by \swift-XRT and \nustar; 3 -- on 2022 October 13 by \swift-XRT; 4 -- 2023 on April 15 by \ixpe, \swift-XRT, NICER, and \nustar and ATCA; 5 -- on 2023 April 16 by \ixpe, NICER, \nustar, and \swift-XRT. From this HID we observe that the source was in the Lower Banana state in all of them. To further confirm this result, we used coordinated \swift-XRT data to monitor the hardness ratio (HR) along the two periods of observations; we defined the \swift-XRT HR as follows:
\begin{equation}
\text{HR} = \frac{ \text{counts in 4--10\,keV} - \text{counts in 0.3--4\,keV}}{\text{counts in 0.3--10\,keV}}.
\end{equation}
Figure \ref{fig:hr} shows this HR as a function of time near the first \ixpe pointing -- when, because of the brightest ever gamma-ray burst \citep{Burns_2023}, the \ixpe observations were stopped, and other observatories were not able to perform strictly simultaneous observations -- and
near the second \ixpe pointing when all the observations with other telescopes were strictly simultaneous.

From this analysis, we can confirm that the HR values fluctuate around an average HR band as shown in Figure \ref{fig:hr} thus the source has not changed its state; only the data from ObsID 00014980055 (on April 15 since 00:41) show a slightly harder spectrum, but this hardening corresponds to a slightly higher counting rate, and it can be explained as due to pile-up \citep{Romano2006}. Also from the \nustar HR, assuming the soft component in 3--10 keV and the harder one in 10--30 keV, we see that HR does not vary much.
Thus, for the following spectral and spectro-polarimetric analysis, we join the data from the different observations to improve the sensitivity.

\begin{figure} 
\centering
\includegraphics[width=0.46\textwidth]{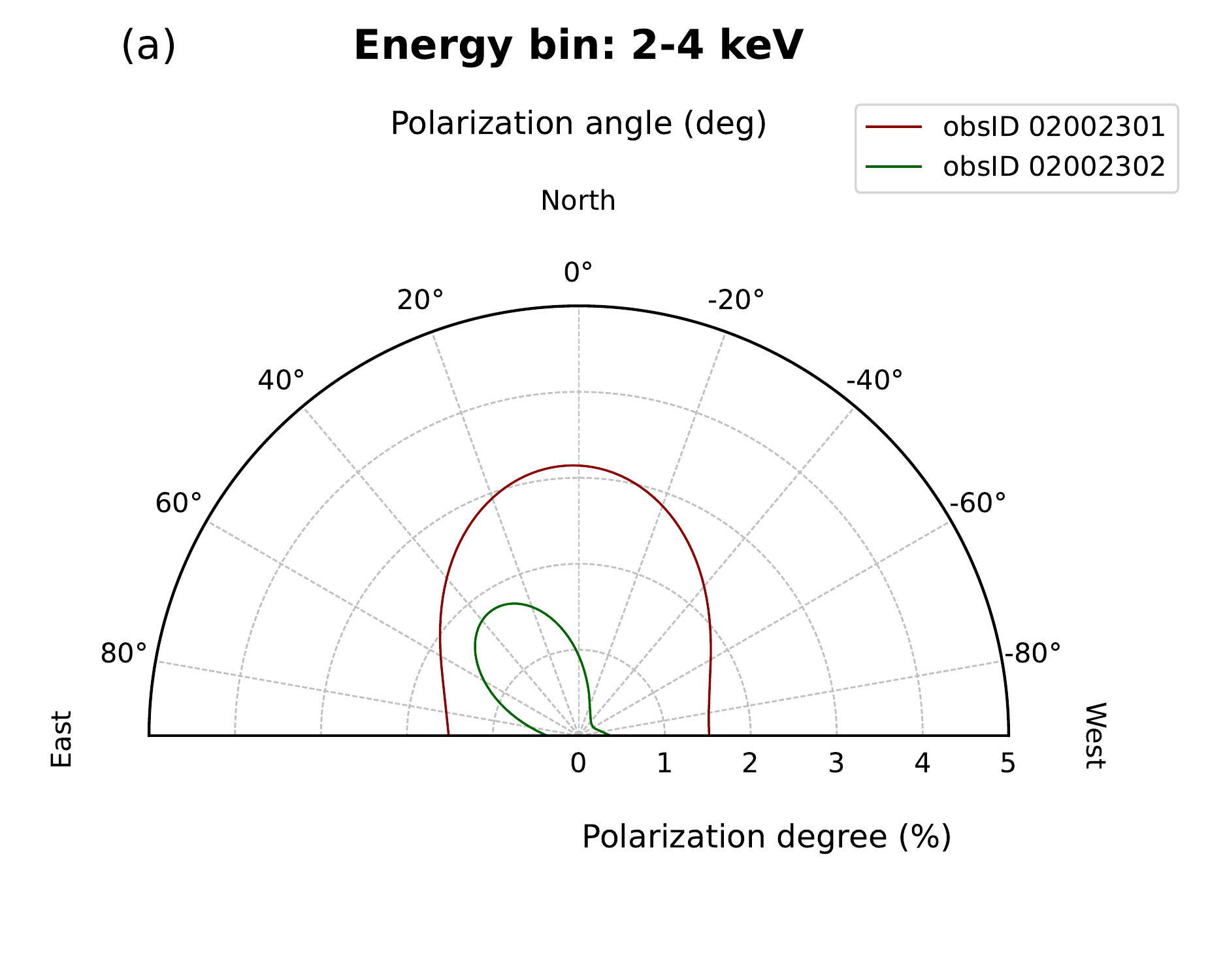}
\includegraphics[width=0.46\textwidth]{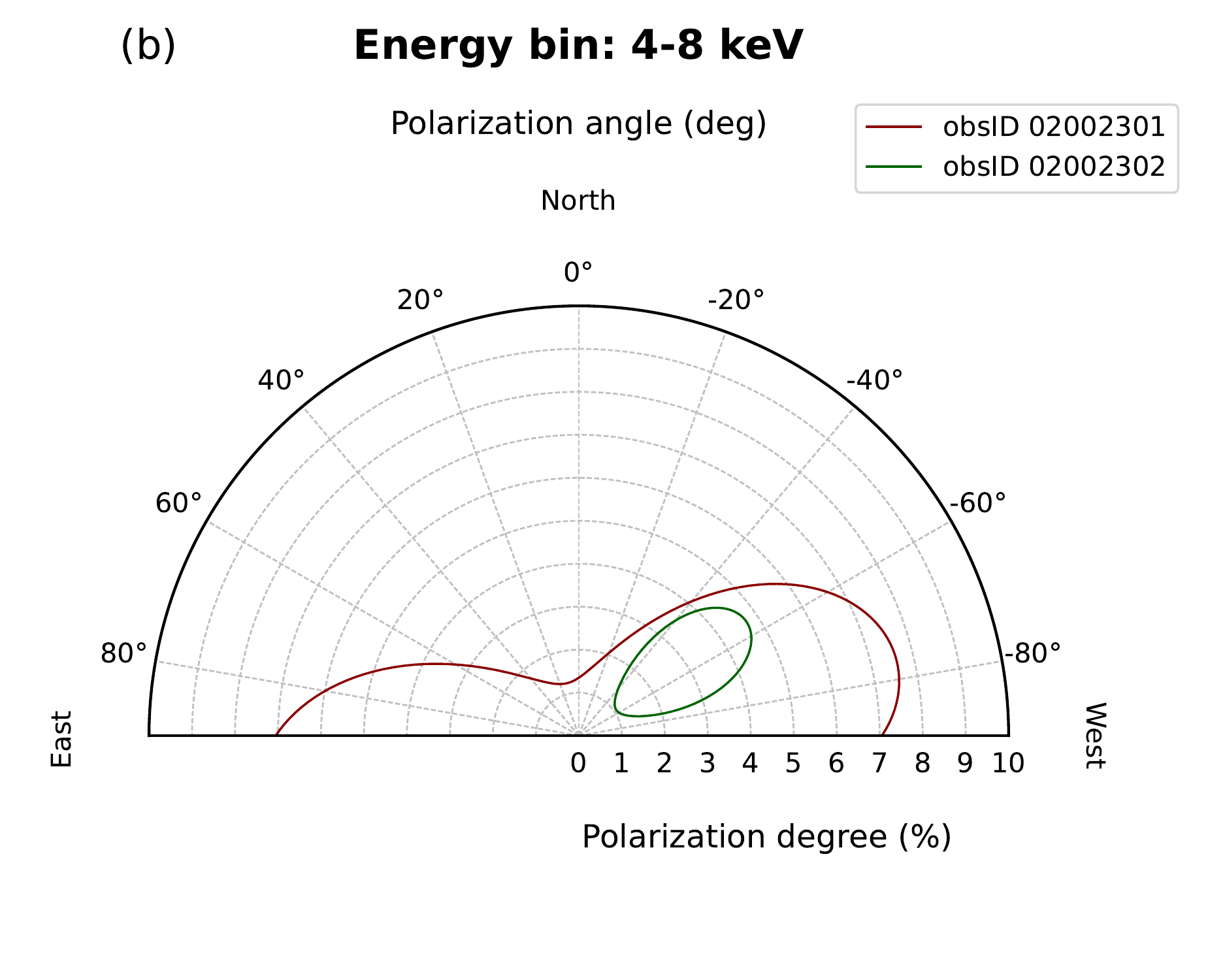}
\includegraphics[width=0.46\textwidth]{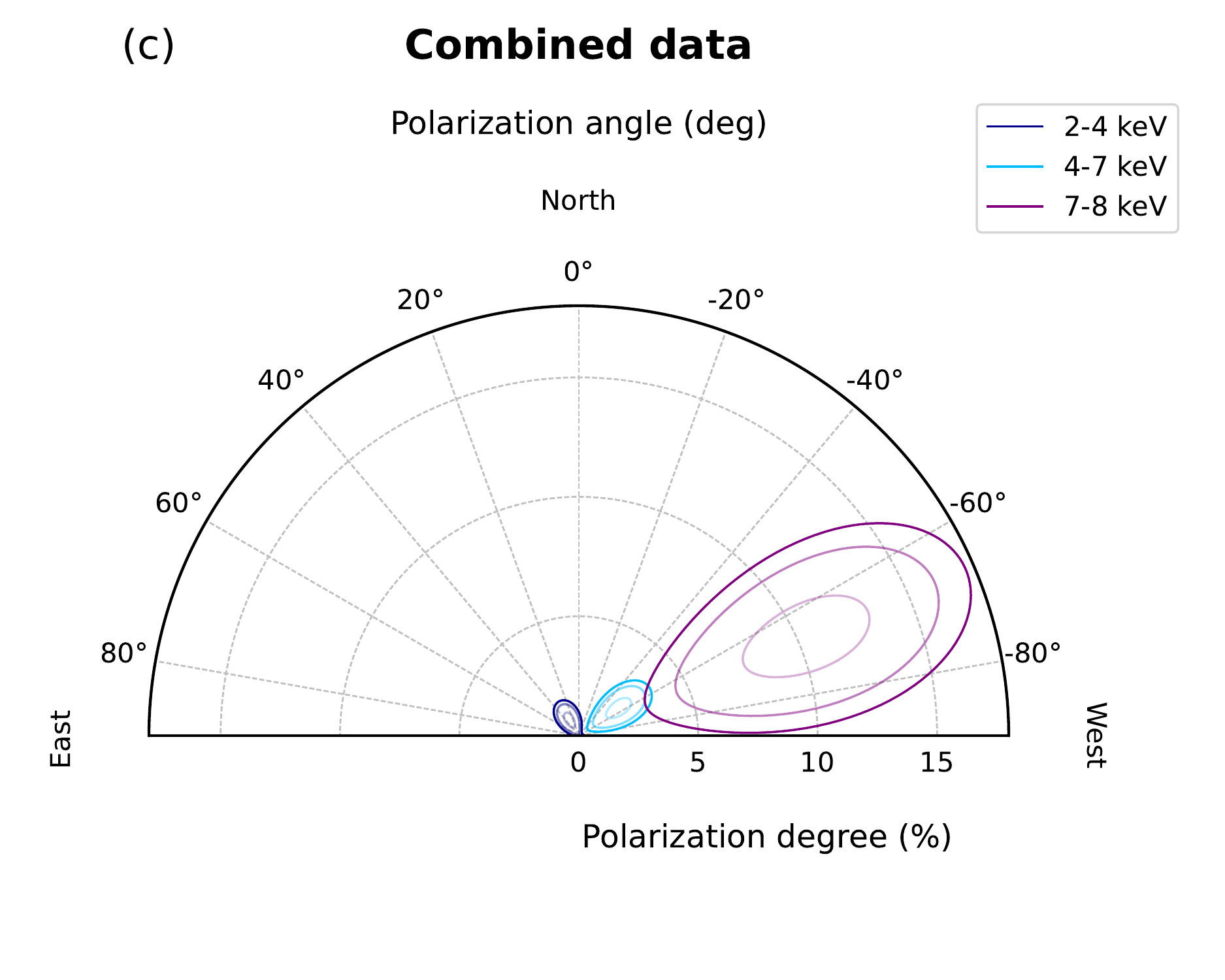}
\caption{Polar plot of the X-ray polarization in 4U~1820$-$303. 
(a) Contours at 99\% CL in the 2--4\,keV energy bin, for the first (red) and the second (green) observations. 
(b) Same as (a), but in the 4--8\,keV energy bin. 
(c) Combined data in the three energy bins 2--4, 4--7, and 7--8\,keV. The contours correspond to 50\%, 95\%, and 99\% CL.}
\label{fig:polar_2bins}
\end{figure}

\section{Polarimetric analysis}

\ixpe observed 4U~1820$-$303 twice, both times during its lower banana state; this suggests that the two observations have the same spectral properties, allowing us to merge them. From the point of view of the polarization, due to the lack of significant measurements before \ixpe, this is something that we need to confirm by comparing the polarization in the first and in the second observation. At this aim, we used the \textsc{ixpeobssim} software \citep{Baldini22} with the \texttt{PCUBE} algorithm to have a preliminary estimate of polarization. In the \ixpe data, the background is negligible (roughly two orders of magnitude below the source in the whole energy band); thus, the prescription reported in \cite{Di_Marco2023} for bright sources have been applied.

\ixpe in its nominal energy band reports an average polarization that is not significant in the first observation and has a low significance in the second. 
Also combining the two data sets, we obtain a polarization below the 99\% CL.
The polarimetric analysis was performed using two energy bins: 2--4 and 4--8\,keV, as reported in Figure~\ref{fig:polar_2bins}(a) and (b). 
The contour regions resulting from the two \ixpe observations in both the energy bins are compatible allowing to combine the two data sets.

\begin{figure} 
\centering\includegraphics[width=0.95\linewidth]{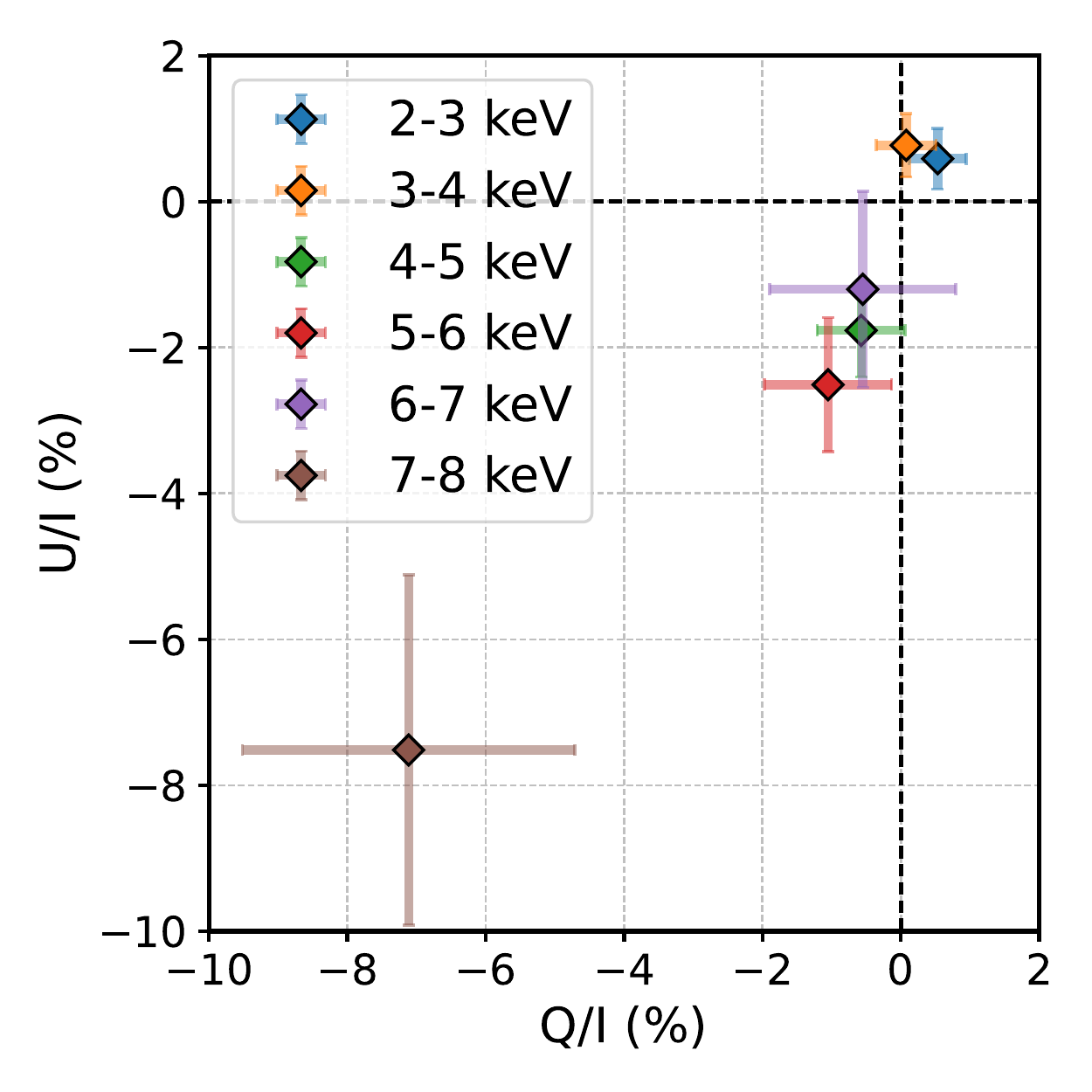} \\
\includegraphics[width=0.95\linewidth]{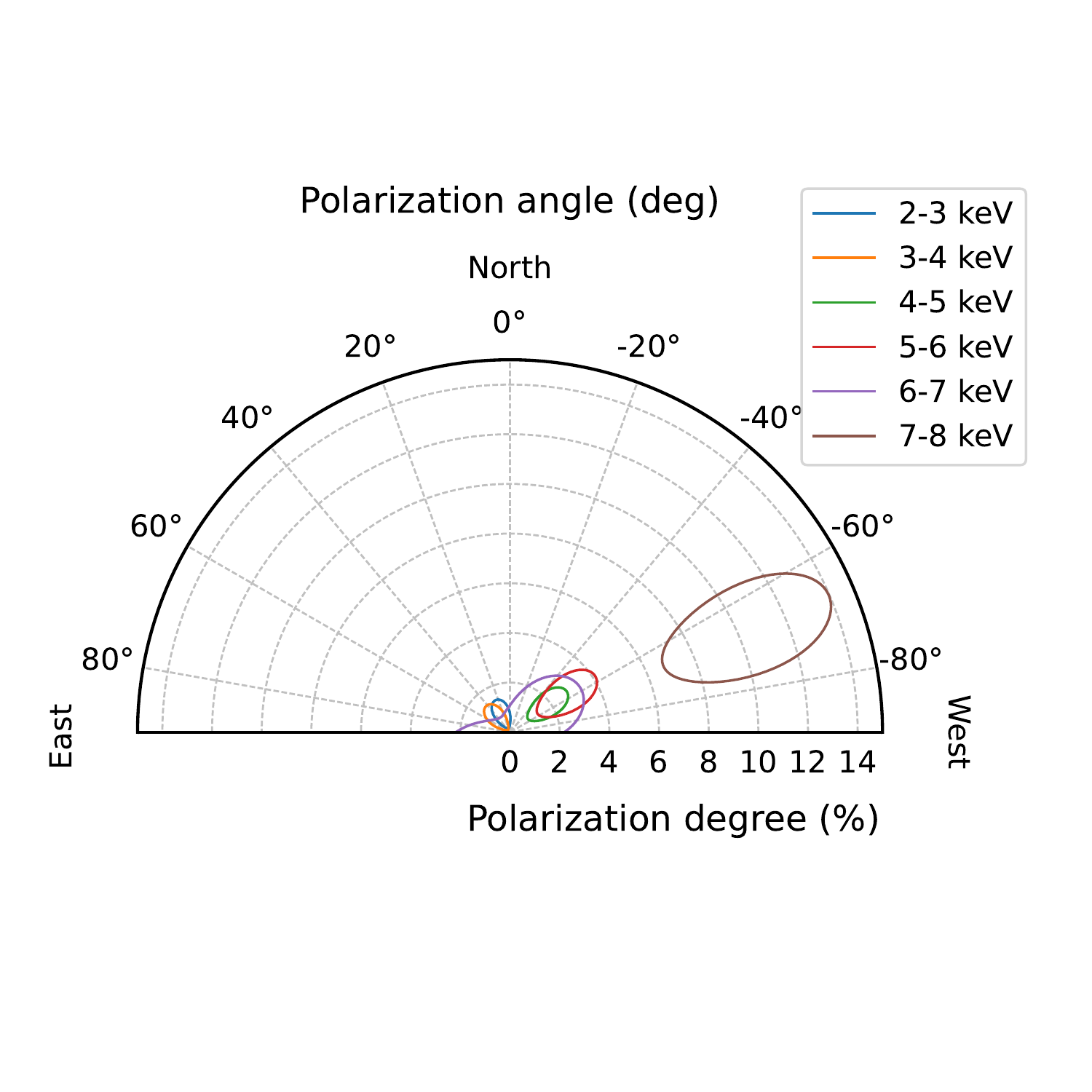}
\caption{Top: normalized Stokes parameters $Q/I$ and $U/I$ in different energy bins as measured with \texttt{PCUBE} in \textsc{ixpeobssim}. Bottom: polar plot of the X-ray polarization in 4U~1820$-$303 when 1 keV binning is applied. Contours are reported at 68\% CL.
A trend of polarization growing with energy is visible in both plots. 
}
\label{fig:stokes_pcube}
\end{figure}

We performed a study of the polarization using the combined data from the two \ixpe observations by using the \texttt{PCUBE} analysis to measure the normalized Stokes parameters as a function of the energy with bins of 1\,keV; the results are shown in Figures~\ref{fig:stokes_pcube}, and reported in Table~\ref{tab:pol_pcube}. We obtain that only the energy bin 7--8 keV has a highly significant polarization at 99.99\% CL. On the other hand, taking into account each energy bin as an independent data set, the
statistical significance of this observation in any one bin has been evaluated; then, the CL for the polarization detection was tested assuming the ensemble of bins against the null hypothesis (i.e. assuming a null polarization in every bin). The result of this test allowed us to obtain a probability of null polarization of 5.32$\times 10^{-6}$, which correspond to a polarization detection at level of 99.995\% CL. Moreover, we see that the Stokes parameters show a clear trend of the polarization degree (PD) increasing with energy. 
We identify three  energy ranges showing similar polarization from Figure~\ref{fig:stokes_pcube}: 2--4, 4--7, and 7--8\,keV. 
By grouping the data within these wider bins, we can increase the significance of detection both in each bin (2--4 keV and 4--7 keV have a polarization detection at 96\% and 99.97\% CL, respectively) and for the whole set of bins reaching a 99.99997\% CL. 
The resulting PD and PA contour plots for the three wider bins are shown in Figure~\ref{fig:polar_2bins}(c), while the numerical results are reported in Table~\ref{tab:pol_pcube}.
The most significant polarization is measured at CL $>99.99$\% in the 7--8\,keV energy bin, where the PD reaches a value of $\sim$10\%, never observed in other weakly magnetized sources \citep{Farinelli23,Capitanio23,Chatterjee2023,Jayasurya2023, Cocchi2023,Ursini2023}. 
We also observe a position-angle change by nearly 90\degr\ between 2--4\,keV, and the higher energy bins (significant at 96\% CL).

\begin{deluxetable}{ccccc}
\tablewidth{0pt}
\tablecaption{Polarization properties obtained with the \texttt{PCUBE} algorithm for the merged data from both the IXPE observations.}
\label{tab:pol_pcube}
\tablehead{
\colhead{Energy bin} & \colhead{PD} & \colhead{PA} & 
\colhead{$Q/I$} & \colhead{$U/I$}  \\ 
\colhead{(keV)}  & \colhead{(\%)} & \colhead{(deg)} & \colhead{(\%)} & \colhead{(\%)}    
}
\startdata
2--3    & $0.8\pm0.4$ & $24\pm15$ & $0.5\pm0.4$ & $0.6\pm0.4$ \\ 
3--4    & $0.8\pm0.4$ & $42\pm16$ & $0.1\pm0.4$ & $0.8\pm0.4$ \\ 
4--5    & $1.9\pm0.6$ & $-54\pm10$ & $-0.6\pm0.6$ & $-1.8\pm0.6$  \\ 
5--6    & $2.7\pm0.9$ & $-56\pm10$ & $-1.1\pm0.9$ & $-2.5\pm0.9$  \\ 
6--7    & $1.3\pm1.3$ & $-57\pm29$ & $-0.6\pm1.3$ & $-1.2\pm1.3$  \\ 
7--8    & $10.3\pm2.4$ & $-67\pm7$ & $-7.1\pm2.4$ & $-7.5\pm2.4$  \\  
\hline
2--4    & $0.75\pm0.30$ & $30\pm11$ & $0.37\pm0.30$ & $0.66\pm0.30$ \\ 
4--7    & $2.0\pm0.5$ & $-55\pm7$ & $-0.7\pm0.5$ & $-1.9\pm0.5$  \\ 
\enddata
\tablecomments{Errors are reported at 68\% CL.}
\end{deluxetable}

%
%

\section{Spectral analysis}\label{sec:spectral_model}

\begin{figure} 
\centering
\includegraphics[width=0.95\linewidth]{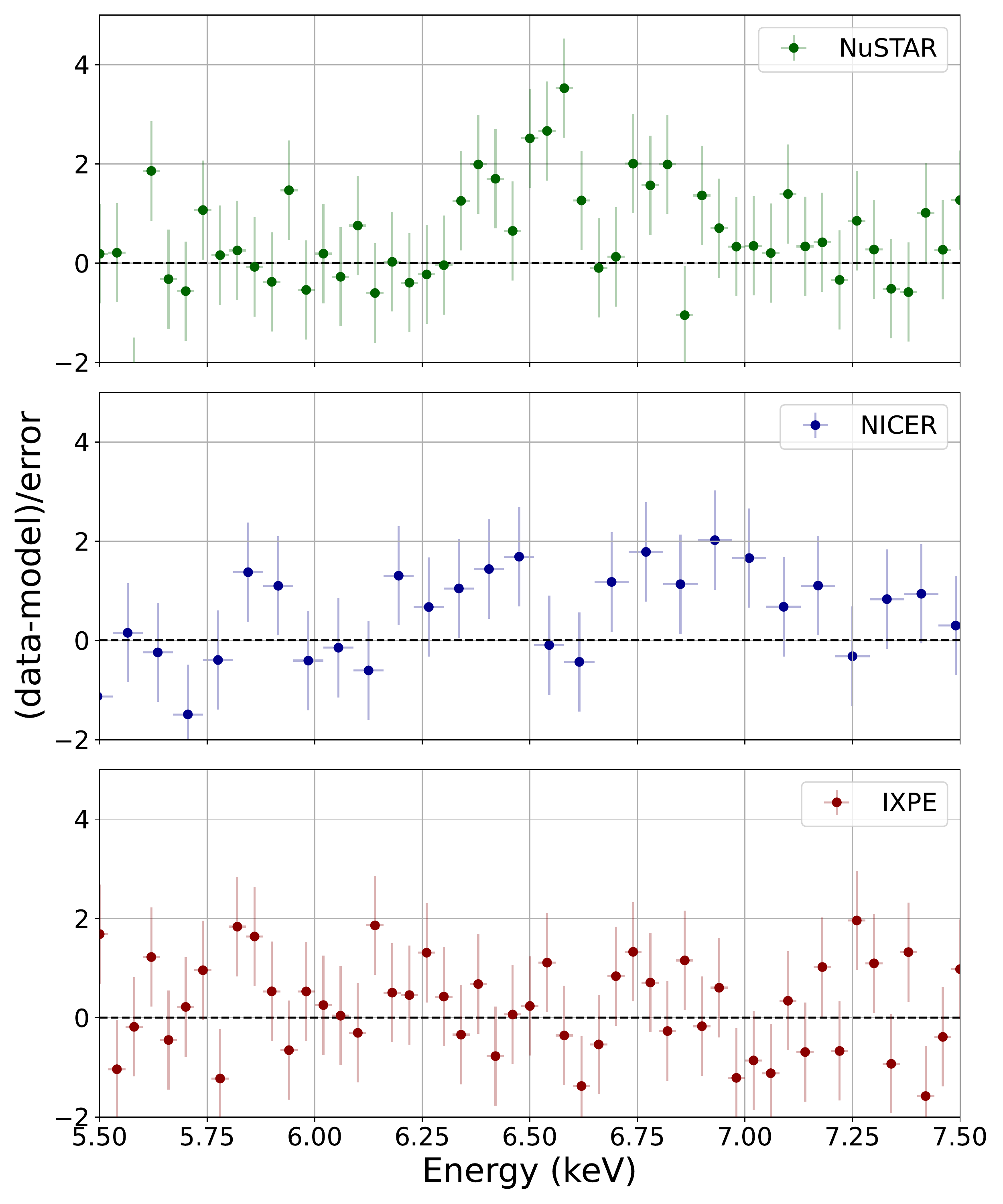}
\caption{Energy spectra residuals in the energy range 5.5--7.5\,keV from the 4U~1820$-$303 observations with \nustar (top), NICER (center), and \ixpe (bottom), when the spectral model does not include the iron line. We see that \nustar data show an excess compatible with a broad iron line, not appreciable in \ixpe and NICER data.}
\label{fig:feline}
\end{figure}

Using the data from NICER (in 0.7--12\,keV) and \nustar (in 3--30\,keV) -- see Appendix~\ref{sec:dataX} for details about data extraction -- and \ixpe (in 2--8\,keV) weighted spectra \citep{DiMarco_2022} and on the basis of the spectral models present in the literature \citep{Titarchuk_2013, Tarana_2007}, we applied the model \texttt{tbabs*(diskbb+comptb)} in \textsc{xspec} v.12.13.0c \citep{Arnaud96}. To estimate the absorption from the interstellar medium, we set the abundances at the \texttt{wilm} values \citep{2000ApJ...542..914W}.
The \nustar residuals show an excess, as reported in Figure~\ref{fig:feline}, compatible with the presence of a broad iron line, such an excess is not appreciable in the \ixpe and NICER data.

Because of the excess in the \nustar residuals, we included in the model a Gaussian line: \texttt{tbabs*(diskbb+comptb+gauss)}. The $\chi^2$/dof is 1955/1785 without a line, to compare with 1835/1780 when a line is included. Summarizing, the adopted best-fit model includes  \texttt{diskbb} for the disk emission, \texttt{gauss} for the Fe K${\alpha}$ line, and \texttt{comptb} (with bulk parameter $\delta = 0$) for thermal Comptonization  with the seed photons following a blackbody distribution ($\Gamma = 3$) and a large Comptonized fraction $A \gg 1$ that is obtained by freezing $\log A = 5$. We tested also \texttt{bbody} for the soft component and the \texttt{compTT} for the harder one, obtaining similar results, but we opted for this spectral model that provides a better reduced $\chi^2$.

\begin{figure} 
\centering
\includegraphics[width=0.9\linewidth]{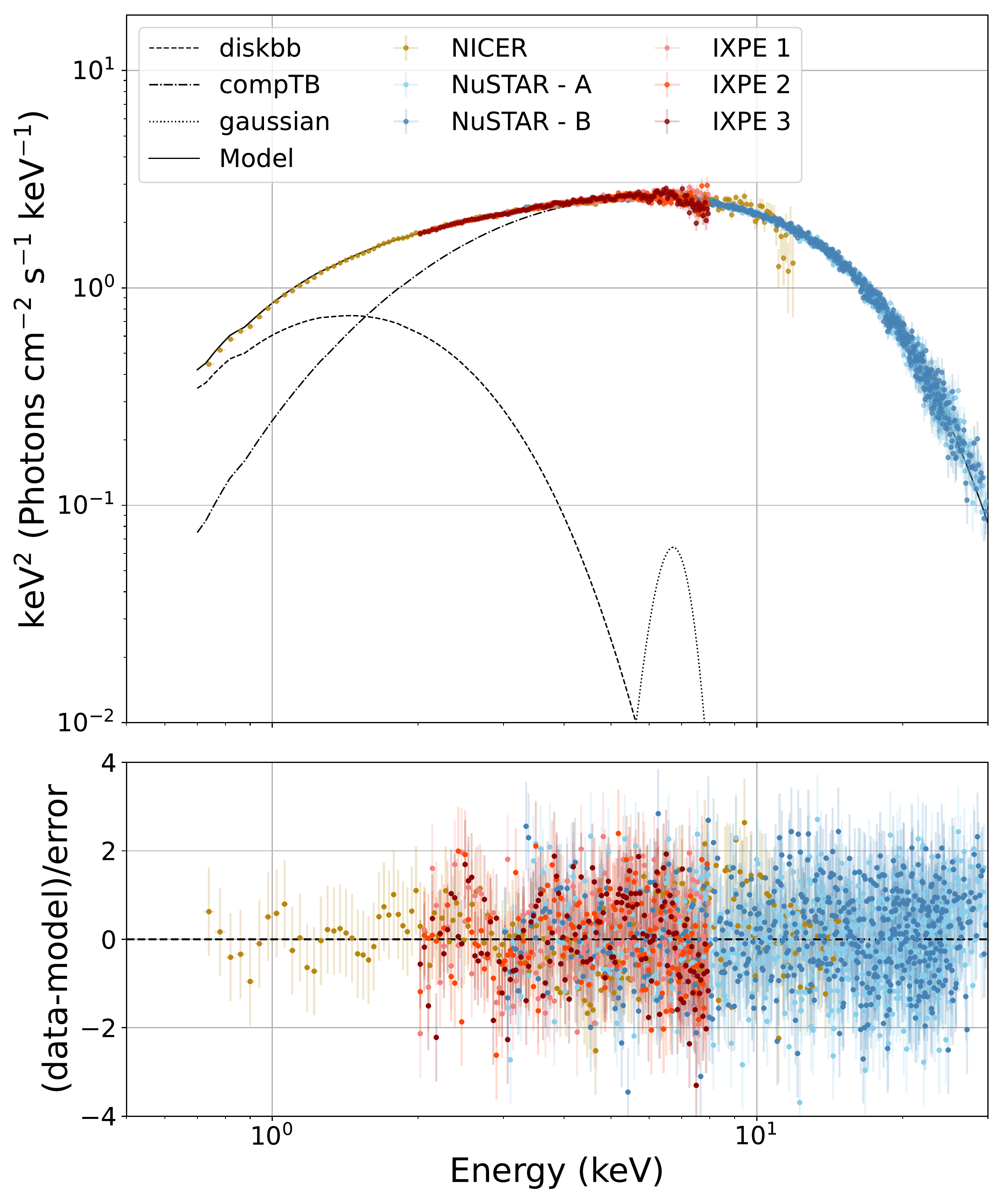}
\caption{Spectral energy distribution of 4U~1820$-$303 in $EF_E$ representation. The points show the data from NICER (brown), \nustar (blue), and \ixpe (red).
The different spectral model components are reported in black lines for \texttt{diskbb} (dashed), \texttt{comptb} (dotted-dashed), and \texttt{gauss} (dotted). 
The bottom panel shows the residuals between the  data and the best-fit model.}
\label{fig:bestfit}
\end{figure}

This model has been applied to fit a broadband spectrum of the combined simultaneous dataset by NICER, \nustar, and \ixpe. 
We also used the energy-independent cross-normalization factors. 
The best-fit parameters are summarized in Table~\ref{tab:spectrum}. The $EF_E$ spectrum and its residuals are reported in Figure~\ref{fig:bestfit}.

\begin{deluxetable}{crl}
\tablecaption{Best-fit parameters of the spectral model \texttt{tbabs(diskbb + comptb + gauss)} applied to the simultaneous data from NICER, \nustar, and \ixpe. The estimated flux in 2--8 keV is $5.23\times 10^{-9}$\flux.}
\label{tab:spectrum}
\tablehead{
\colhead{Model} &
\colhead{Parameter} &
\colhead{Value}  
}
\startdata
    \texttt{tbabs} & $N_{\rm H}$ ($\times 10^{22}$ cm$^{-2}$) & $0.14\pm0.01$ \\
    \hline
    \texttt{diskbb} & $kT_{\rm in}$ (keV) & $0.52^{+0.03}_{-0.04}$ \\
    & \text{norm} ($[R_{\rm in}/D_{10}]^2\cos\theta$) & $1900^{+500}_{-400}$ \\
    & $R_{\rm in}$\tablenotemark{a} (km) & $40\pm20$ \\
    \hline
    \texttt{comptb} & $kT_{\rm s}$ (keV) & $0.72\pm0.05$ \\
    & $\alpha$ & $0.93\pm0.01$\\
    & $kT_{\rm e}$ (keV) & $2.919^{+0.016}_{-0.008}$\\
    & $L_{\rm X}$ \tablenotemark{b} ($\times 10^{39}$ \lum) & $0.0347 \pm 0.0004$ 
    \\ \hline
    \texttt{gauss} & $E_{\rm line}$ (keV) & $6.63\pm0.09$ \\
    & $\sigma$ (keV) & $0.56\pm 0.01$ \\
    & \text{norm} (photon~cm$^{-2}$~s$^{-1}$) & $0.0020^{+0.0006}_{-0.0005}$ \\
    & Equivalent width (eV) & 34.2$\pm$1.2\\ 
    \hline
    \multicolumn{3}{c}{$\chi^2$/dof = 1835/1780 = 1.03} \\
    \hline
    \multicolumn{3}{c}{Cross normalization factors} \\
   &  $C_{\rm NICER}$ & 1.0 \\
    & $C_{\rm \nustar-A}$ & $1.063\pm0.004$ \\
   & $C_{\rm \nustar-B}$ & $1.057\pm0.004$ \\
   &  $C_{\rm \ixpe-DU1}$ & $1.034\pm0.007$ \\
   & $C_{\rm \ixpe-DU2}$ & $0.990\pm0.007$ \\
   & $C_{\rm \ixpe-DU3}$ & $0.932\pm0.007$ \\
    \hline
    \multicolumn{3}{c}{Photon flux ratios in 2--8\,keV} \\
   & $F_{\rm diskbb}/F_{\rm tot}$ & 0.114 \\
  & $F_{\rm comptb}/F_{\rm tot}$ & 0.884 \\
   & $F_{\rm gauss}/F_{\rm tot}$ & 0.002 \\
\enddata
\tablecomments{Errors are reported at 90\% CL.}
\tablenotetext{a}{The inner radius for the \texttt{diskbb} component is estimated assuming an inclination at 40\degr, as reported in \citet{Anderson_1997}, and a distance of 8~kpc \citep{2021MNRAS.505.5957B}.}
\tablenotetext{b}{The source luminosity for $D=8$~kpc.}  
\end{deluxetable}
To take into account \nustar calibration uncertainties \citep{madsen22}, we assumed a gain offset free in the fit, obtaining $-(0.085\pm0.009)$\,keV and $-(0.068\pm0.008)$\,keV for the focal plane modules A and B (FPMA and FPMB), respectively; similarly for \ixpe calibration uncertainties \citep{DiMarco_2022b} we left free the gain slope and offset obtaining a slope of $0.981 \pm 0.003$\,keV$^{-1}$, $0.973 \pm 0.003$\,keV$^{-1}$, $0.980 \pm 0.003$\,keV$^{-1}$ for the DU1, DU2 and DU3 respectively, while the offset in each one is $0.003 \pm 0.012$\,keV, $0.032 \pm 0.012$\,keV and $0.020 \pm 0.012$\,keV. As reported also by the photons fluxes of Table~\ref{tab:spectrum}, the spectrum in the whole 2--8\,keV \ixpe energy band is dominated by the Comptonization component, while the disk contributes only at lower energies (see also Figure~\ref{fig:bestfit}). 

It is worth noting that the Fe line is typically found along with a broadband reflection component \citep[e.g.,][]{2016A&A...596A..21I,Ursini2023}. We therefore tested several self-consistent models (such as \texttt{relxillns}) or convolution ones (such as \texttt{rdblur*rfxconv})
to take into account the reflection component. However, none of them produced a statistically significant improvement to the fit, and the reflection fraction was negligible ($<$5\%). Since a more detailed treatment of the reflection goes beyond the scope of this manuscript, we did not include any reflection component in the following spectro-polarimetric modeling.

\begin{deluxetable*}{rrccc} 
\tablecaption{Main results of the spectro-polarimetric analysis}
\label{tab:specpol}
\tablehead{ \colhead{Polarimetric} & & \colhead{\texttt{polconst*diskbb +}}  & \colhead{\texttt{polconst*diskbb +}} & \colhead{\texttt{polconst*diskbb +}} \\
\colhead{components} & & \colhead{\texttt{polconst*comptb}} & \colhead{\texttt{pollin*comptb}} & \colhead{\texttt{polpow*comptb}} 
 }
\startdata
\texttt{diskbb} & PD (\%) & 9.8$\pm4.2$ & 8.1$^{+7.1}_{-6.8}$ & 3.2$^{+3.0}_{-2.9}$ \\
    & PA ($\deg$) & $32 \pm 13$ & $-59^{+13}_{-26}$ & $43 \pm 35$ \\ \hline
\texttt{comptb} & PD / $A_1$  (\%)\tablenotemark{a} & $5.31 \pm 0.24$ & $5.7\pm2.4$ & $0.46^{+9.63}_{-0.46}\times10^{-3}$ \\
    & $A_{\rm slope}$ (\%\,keV$^{-1}$) & -- & $-1.9\pm0.6$ & -- \\
    & $A_{\rm index}$  & -- & -- & $-4.9^{+1.6}_{-2.6}$ \\
    & PA / $\psi_1$  (deg)\tablenotemark{a} & $-63\pm 11$ & $38\pm7$ & $-63\pm7$ \\
    & $\psi_{\rm slope}$ (deg\,keV$^{-1}$) & -- & $0$ (f)\tablenotemark{b} & -- \\
    & $\psi_{\rm index}$  & -- & -- & $0$ (f)\tablenotemark{b} \\\hline
    & $\chi^2$/dof & 724/662 = 1.094 & 699/661=1.057 & 697/661=1.054\\ 
\enddata
\tablecomments{Errors are at 90\% CL. }
\tablenotetext{a}{For the \texttt{pollin} and \texttt{polpow} models, $A_1$ and $\psi_1$ refer to 
the PD and PA values at 1\,keV.}
\tablenotetext{b}{We fixed the slope and the index of the PA at zero.}
\end{deluxetable*}

\section{Spectro-polarimetric analysis}

The \ixpe Stokes parameters $I$, $Q$, and $U$ spectra have been fitted with \textsc{xspec} v.12.13.0c \citep{Arnaud96} freezing the spectral model at the one reported in Sect.~\ref{sec:spectral_model} and summarized in Table~\ref{tab:spectrum}, and applying the same gain corrections to the response files of the $I$, $Q$, and $U$ spectra. 
As discussed above, we tested different models, but for the spectro-polarimetric analysis in this section, we opt for the simplest model satisfactorily fitting the data, which is \texttt{tbabs*(diskbb+gauss+comptb)}.

\begin{figure*} 
\centering
\includegraphics[width=0.85\textwidth]{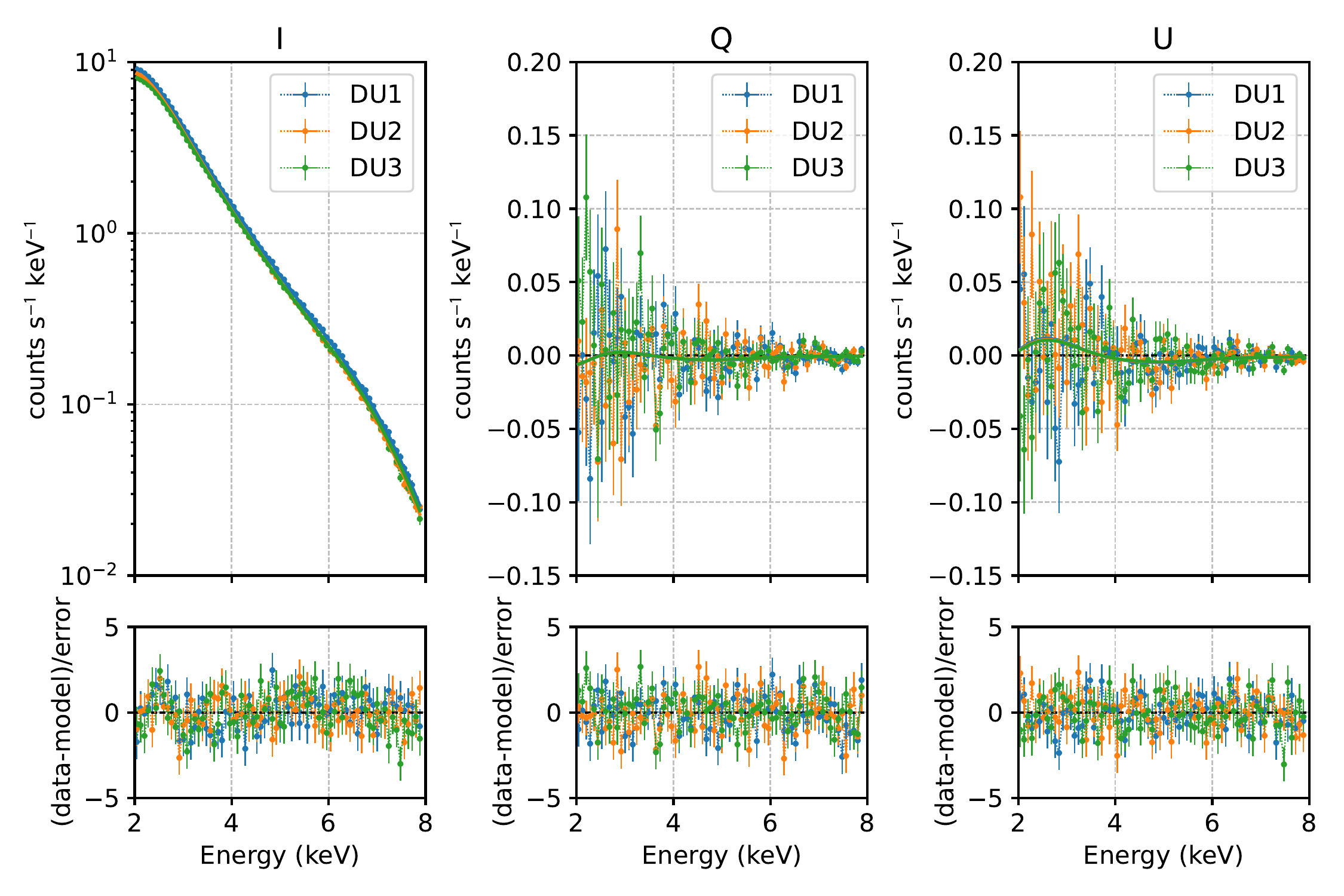}
\caption{Spectral joint fit for the Stokes parameters $I$, $Q$, and $U$ in the 2--8 keV energy band using three IXPE detectors and applying the model \texttt{tbabs*(gauss+polconst*diskbb+polpow*comptb)}. The best-fit gives $\chi^2$/dof =1.054 (see also Table \ref{tab:specpol}).}
\label{fig:spec_pol}
\end{figure*}

The $I$ spectrum alone gives a $\chi^2$/dof = 226/212 = 1.07. 
The first spectro-polarimetric analysis, including $Q$ and $U$ spectra, has been performed to confirm an increase of the polarization with energy; thus, we tested two models, the first one assuming a constant polarization using the \texttt{polconst} model from \textsc{xspec} for both the \texttt{diskbb} and \texttt{comptb} components, obtaining a PD of 0.5\%$\pm$0.3\%, PA =$-75\degr\pm27\degr$, and $\chi^2$/dof = 740/664 = 1.11. Hereafter, all the errors are reported at 90\% CL.
Then, we considered the possibility that polarization linearly depends on energy using the \texttt{pollin} model: $\mbox{PD}(E)=A_1+A_{\rm slope}(E-1)$, $\mbox{PA}(E)=\psi_1+\psi_{\rm slope}(E-1)$ (where the energy $E$ is in keV). 
We obtain the PD at 1\,keV of $A_1=2.9\%\pm1.0\%$ with a slope of $A_{\rm slope}=-1.3\%\pm0.3$\%~keV$^{-1}$ and a PA of $\psi_1=43\degr\pm27\degr$ at 1 keV with a slope $\psi_{\rm slope}=-3\degr\pm5\degr$~keV$^{-1}$, with a $\chi^2$/dof = 704/662 = 1.063. 
These results are compatible with the PA being constant, while the PD has a negative slope. 
We note that the best-fit PD at 1~keV and the PD slope imply that the PD becomes close to zero at about 3~keV and reaches a value of $\sim$6\% at 7--8 keV at a PA of $-47\degr$ (change of sign of PD is equivalent to a 90\degr\ swing of the PA).  
This confirms the results obtained with \texttt{PCUBE} shown in Figures~\ref{fig:stokes_pcube} and \ref{fig:polar_2bins}(c). 
The F-test for these two fits gives a value of 17.13 (probability 5.6$\times10^{-8}$), confirming that the energy-dependent polarization is favored by the data. 

Considering that the PD at low energy is not significant (see Figure~\ref{fig:polar_2bins}(c) and Table~\ref{tab:pol_pcube}), we tested the hypothesis that only the \texttt{comptb} component is constantly polarized, but in this scenario, $\chi^2$/dof reaches a value higher than 4, confirming the need for a polarized \texttt{diskbb} component in the spectro-polarimetric fit.
At this point, we tried to disentangle the polarization of the \texttt{diskbb} and \texttt{comptb} components by associating each component with its own polarization. 
We assumed for the disk a constant polarization, and for the Comptonization, we considered three different polarimetric models: \texttt{polconst} with a constant polarization, \texttt{pollin}, i.e., polarization depends linearly on energy,  and \texttt{polpow} model that assumes that polarization is changing as a power law of energy $\mbox{PD}(E)=A_{\rm norm}E^{-A_{\rm index}}$ and $\mbox{PA}(E)=\psi_{\rm norm}E^{-\psi_{\rm index}}$.  
In order to reduce the number of parameters, we fixed $\psi_{\rm slope}$ and  $\psi_{\rm index}$ at zero.
The results from this analysis are reported in Table~\ref{tab:specpol} and Figure~\ref{fig:spec_pol}.

All considered models give similar results, with polarization at low energies being orthogonal to that at higher energies. 
For the \texttt{polconst} model, the disk polarization reaches $\sim$10\%, the Comptonization component is significantly polarized at a 5\% level, with the PA being orthogonal to that of the disk.  
In the case of the \texttt{pollin} model for the \texttt{comptb} component, the disk polarization becomes compatible with smaller values, as expected in literature, while, at the same time, the PD of the Comptonization component changes sign around 4~keV corresponding to the rotation of the PA from 28\degr\ at lower energies to $-62\degr$ at higher energies confirming the result of \texttt{PCUBE}.  
For the \texttt{polpow} model, we obtain a PD strongly increasing with energy (note the negative PD index $A_{\rm index}\approx-5$) from a PD at 1\,keV of just $\sim10^{-3}$\%.   
The uncertainties on the parameters are quite large due to the correlation between PD and PA, but also because the statistical uncertainty of the data does not allow us to clearly disentangle different models. 

\section{Radio observation} \label{sec:radio}

4U~1820$-$303 was observed with the Australia Telescope Compact Array (ATCA) on 2021 April 15 from 12:49:40 and 21:51:20 UT (under project code CX530). During this observation, ATCA was in a relatively compact H214 configuration. The data were recorded simultaneously at two central frequencies, 5.5 and 9\,GHz, with 2\,GHz of bandwidth at each frequency. We used the unpolarized PKS~B1934$-$638 for primary calibration and to solve for antenna leakages. The nearby calibrator B1817$-$254 was used for gain calibration. The calibration and imaging followed standard procedures using the Common Astronomy Software Applications for radio astronomy (\textsc{casa}, version 5.1.2; \citealt{2022PASP..134k4501C}). The polarization calibration used the \textsc{casa} task \texttt{atcapolhelpers.py} and \texttt{qufromgain} routine.\footnote{From {\url{https://github.com/radio-astro}}} The imaging used a Briggs robust parameter of 2 to maximize the image sensitivity. The isolated antenna 6 (located 6\,km from the array core) was used during imaging.

While 4U~1820$-$303 was detected at both frequency bands, the radio counterpart was relatively faint. Fitting for a point source in the image plane (Stokes $I$), we measured a flux density of $100 \pm 12\,\mu$Jy at 5.5\,GHz and $80 \pm 10\,\mu$Jy at 9\,GHz. This corresponds to a radio spectral index of $-0.45 \pm 0.30$. The X-ray brightness, state, and radio spectral index at the time of the observations suggest that the radio emission originates from either a quenched compact jet or transient ejecta (see, e.g., \citealt{2021MNRAS.508L...6R}). Imaging the field in both Stokes $Q$ and $U$, no significant linear polarization was detected.
We place a 3$\sigma$ upper limits on the fractional linear polarization of 60\% and 70\% at both 5.5 and 9\,GHz, respectively. Stacking the two bands gives a 3$\sigma$  upper limit of 50\% (at 7.25\,GHz). 

\section{Discussion}

In this paper, we report the first detection of polarization in the X-rays with \ixpe and a new constraint on radio polarization by ATCA for the atoll source 4U~1820$-$303. 
The spectral analysis performed using  the data from different X-rays observatories confirms the presence of a broad Fe line, as reported in \citet{Cackett_2008}, \citet{Titarchuk_2013}, and \citet{Mondal2016}, in the NuSTAR data. In the radio, the source showed an emission that is consistent with either a quenched compact jet or a transient ejecta. 

\ixpe results on the X-ray polarization obtained  using the model-independent \textsc{ixpeobssim}--\texttt{PCUBE} analysis show a different behavior of this source with respect to the other atoll and Z-sources observed up to now \citep{Farinelli23,Capitanio23,Chatterjee2023,Jayasurya2023, Cocchi2023, Ursini2023}. 
In particular, this is the first source showing a strong increase of the polarization with energy, up to 10\% in the 7--8\,keV energy band. 
At low energy, there is no secure detection of polarization, even if there is a hint of a $\sim$0.8\% PD at 96\% CL, with a PA orthogonal to the ones at higher energies. 

We also attempted a spectro-polarimetric analysis of the source emission. 
Assuming simple polarimetric phenomenological models for the Comptonization component, a need for non-zero polarization of the disk emission (which dominates in the lower part of the \ixpe band) emerged to account for the overall low PD. 
In this scenario, leaving the parameters unconstrained, the polarization of the disk tends to be orthogonal to the one associated with the Comptonization. The PD varies, depending on the polarimetric models, from a few percent up to $\sim$10\%. 
In the lower boundary, these values are compatible with the polarization expected in case of an electron-scattering dominated optically thick accretion disk, which is below 2\% \citep{chandrasekhar1960, Sobolev63} considering 4U~1820$-$303 inclination -- ranging from $\sim 20\degr$ to 55\degr\ \citep{Cackett_2008, Mondal2016, Anderson_1997}. 
We also note that, in case of a disk whose opacity is dominated by electron scattering, the PA is expected to be orthogonal to the disk axis. 
Therefore, the PA of the disk component measured by \ixpe, which is $\sim30\degr$ East of North, would indicate that the position angle of the disk rotation axis is $\simeq -60\degr$ (or 120\degr).

The Comptonized component is firmly associated with a high and energy dependent polarization. 
For an optically thin corona, the PD can reach 10\%--20\% in slab geometry \citep{PS96,Schnittman2010,gnarini2022,Poutanen23}. 
However, the Thomson optical depth $\tau_{\rm T}$, required to produce the observed spectrum with the photon index of $\Gamma\approx 1.9$ and electron temperature $T_{\rm e}$ of $\sim$3~keV, exceeds 10. 
For such a high optical depth, the PD is even smaller than that given by the classical results of \citet{chandrasekhar1960} and \citet{Sobolev63} for $\tau_{\rm T}\gg 1$ \citep[see Figure~5 in][]{st85}. 
Thus, the $\sim$10\% polarization observed by \ixpe at highest energies is difficult to explain for any reasonable inclination.
Therefore, our results require some non-standard coronal geometry.  

One of the possibilities is the presence of a mildly relativistic outflow from the inner part of the  accretion flow where Comptonization takes place. 
Because of relativistic aberration, the PD at a given inclination becomes larger than that in the static corona with the PA parallel to the disk normal \citep{Beloborodov98,Poutanen23}. 
For an optically thin outflow, the PD of the scattered component can reach 15\%, but the PD of the total emission will be reduced by unscattered radiation. 
Thus, we are forced to assume that the outflow is optically thick; in this case, the total PD can reach 10\%--15\% even at moderate inclinations of 40\degr--60\degr\ \citep[with the PA being still parallel to the disk normal, see Figure~4 in][]{Beloborodov98}. 
The remaining question is why the PD is high only in the 7--8 keV band, but not below. 

An alternative to the outflow would be the presence of a strong reflection already claimed to be present in previous weakly magnetized neutron stars observed by \ixpe \citep{Farinelli23,gnarini2022,Cocchi2023,Ursini2023}. 
As already discussed in Sect.~\ref{sec:spectral_model}, such a component was investigated, and its relative flux in the \ixpe band is estimated to be below 5\% even above 4 keV. 
Because a typical PD for the reflected component is about 20\% \citep{Matt93,Poutanen96}, it 
cannot alone easily explain the observed PD of 10\% above 7~keV and its fast rise with energy. 
Notwithstanding, adding a polarized contribution from the reflection component aligned with that of the corona could partially reduce the PD of the Comptonized component to a level more easily explained with standard geometries. 

Overall, the spectro-polarimetric modeling support a geometry having a SL perpendicular to the disk plane, already described in the context of previous \ixpe observations of atoll sources, with the important difference of the PD for the Comptonized component which is about 10\% for 4U~1820$-$303 instead of a non-detection for GS~1826$-$238 \citep{Capitanio23}, $\simeq 3$\% for GX~9+9 and Cyg~X-2 in the energy band 4--8 keV \citep{Farinelli23,Ursini2023} and $\simeq 5$\% for XTE~J1701$-$462 in the 4--8 keV band in the horizontal branch. 
This can be contrasted to a $\simeq$50\% of Comptonized fraction in GS~1826$-$238 \citep{Capitanio23} and $<70$\% in GX~9+9, where the reflection component is well determined \citep{Ursini2023}. 
This might imply a correlation between the corona dominated spectrum and the high PD, but further observations are needed to disentangle the polarization associated with each spectral component.

Linearly polarized radio emission was not detected, with an upper limit at 99.73\% CL of 50\% on the PD in the radio band; unfortunately, this result does not allow us to do a direct comparison between the PA in the radio and the X-rays. 
As such, due to the radio faintness, we cannot compare our result to the one obtained for Cyg~X-2 \citep{Farinelli23} and Sco~X-1 \citep{Long1979,Long2022}, where the X-ray PA was aligned with the direction of the radio jets.

\section{Summary}

The \ixpe result for 4U~1820$-$303 shows a different polarization behavior with respect to  other atoll sources, GX~9+9 \citep{Ursini2023,Chatterjee2023} and GS~1826$-$238 \citep{Capitanio23}, but also with respect to the Z-sources Cyg X-2 \citep{Farinelli23} and XTE J1701$-$462 \citep{Jayasurya2023,Cocchi2023}. 
In particular, the need to have a polarized disk emission is more evident.
Moreover, the model-independent analysis gives an indication of a 90\degr\ rotation of the PA at lowest energies, where the disk contributes the most.  
A very high PD of about 10\% detected above 7~keV did not find an obvious explanation within standard  models of X-ray emission of weakly magnetized NSs.  
One possibility is that the polarization is produced in an optically thick outflow emanating from the inner part of the accretion disk. 
Future studies of this source with a longer exposure could help to disentangle  the polarization of each spectral component. 
Monitoring the source along its superorbital period or when it moves along the banana and island states could help to better understand the polarization variations and, thanks to this, verify possible geometry variations as a function of the mass accretion rate.


\section*{Acknowledgments}

The Imaging X-ray Polarimetry Explorer (\ixpe) is a joint US and Italian mission.  The US contribution is supported by the National Aeronautics and Space Administration (NASA) and led and managed by its Marshall Space Flight Center (MSFC), with industry partner Ball Aerospace (contract NNM15AA18C). The Italian contribution is supported by the Italian Space Agency (Agenzia Spaziale Italiana, ASI) through contract ASI-OHBI-2017-12-I.0, agreements ASI-INAF-2017-12-H0 and ASI-INFN-2017.13-H0, and its Space Science Data Center (SSDC) with agreements ASI-INAF-2022-14-HH.0 and ASI-INFN 2021-43-HH.0, and by the Istituto Nazionale di Astrofisica (INAF) and the Istituto Nazionale di Fisica Nucleare (INFN) in Italy.  This research used data products provided by the \ixpe Team (MSFC, SSDC, INAF, and INFN) and distributed with additional software tools by the High-Energy Astrophysics Science Archive Research Center (HEASARC), at NASA Goddard Space Flight Center (GSFC). 

This research has made use of the MAXI data provided by RIKEN, JAXA, and the MAXI team and of the Swift/BAT transient monitor results provided by the Swift/BAT team. We thank Jamie Stevens and ATCA staff for the scheduling of these observations. ATCA is part of the Australia Telescope National Facility (https://ror.org/05qajvd42), which is funded by the Australian Government for operation as a National Facility managed by CSIRO. We acknowledge the Gomeroi people as the Traditional Owners of the ATCA observatory site.

The authors acknowledge the prompt schedules of the simultaneous observations: Keith G. Gendrau, Zaven Arzoumanian, and the NICER SOC team; Boris Sbarufatti, Kim Page, Brad Cenko, and Swift-XRT SOC team;  Karl Forster,   Murray Brightman, Fiona A. Harrison and \nustar SOC team.
We also acknowledge support from the Academy of Finland grants 333112, 349144, 349373, and 349906 (JP, SST) and the German Academic Exchange Service (DAAD) travel grant 57525212 (VD). 

%

\vspace{5mm}
\facilities{\ixpe, Swift--XRT, NICER, NuSTAR}


\software{\textsc{ixpeobssim} \citep{Baldini22}, \textsc{xspec} \citep{Arnaud96}, \textsc{HEASoft} \citep{1995ASPC...77..367B}}

\appendix


\section{Data handling for X-ray observatories}\label{sec:dataX}

In this study, as we reported in the previous sections, \ixpe observations were coordinated with \nustar, NICER, and \swift-XRT aiming to verify the state of the source and also to obtain a better constraint on the spectral model thanks to better energy resolution and broadband coverage. In this section, we briefly report the data handling and extraction applied to the data we used.

\subsection{\ixpe}

The Imaging X-ray Polarimetry Explorer (\ixpe), a NASA mission in partnership with the Italian Space Agency, was launched on 2021 December 9. A detailed description of the observatory and its performance is given in \citet{Weisskopf2022}. \ixpe consists of three identical grazing incidence telescopes, providing imaging and spectral polarimetry over the 2--8\,keV energy band with a time resolution better than 10~$\mu$s. Each telescope comprises an X-ray module of mirror assembly (MMA) and a polarization-sensitive detector unit (DU) equipped with a gas-pixel detector (GPD) \citep{Costa2001, Soffitta_2021}. 

\ixpe observed 4U~1820$-$303 twice, in October 2022 for a short period and for a longer one in April 2023 (ObsID 02002399) with a total effective exposure of 16\,ks and 86\,ks per DU, respectively. Level 2 data were processed with the \ixpe dedicated software \textsc{ixpeobssim} \citep{Baldini22} v. 30.3.0 and with \textsc{ftool} released in HEASOFT v. 6.31.1 using the Calibration database released on 2022 November 17. 

Source photons were selected from the \ixpe telescopes images collected within a circular region having radius 100\arcsec\ centered on the source position, while the background, following the prescription by \cite{Di_Marco2023} for the  sources with high count rate ($\gtrsim2$~cnt s$^{-1}$), has not been subtracted from the data, because it is negligible (in these observations, the background is roughly two orders of magnitude below the source in all energy bins).

For the spectral analysis, the flux (Stokes parameter $I$) energy spectra from the three DUs have been extracted following the weighted approach described in \cite{DiMarco_2022} and data were binned to have at least 30 counts per energy channel, while in the spectro-polarimetric analysis, we applied a constant energy binning of 200 eV for all the three energy distributions of Stokes parameters $I$, $Q$, and $U$. 

\subsection{NICER}

The Neutron Star Interior Composition Explorer (NICER) is a soft X-ray instrument onboard the International Space Station (ISS), launched in June 2017, and it consists of 56 co-aligned concentrator X-ray optics, each of which is paired with a single silicon drift detector. NICER does not offer imaging capabilities, but offers a large collecting area providing unmatched time resolution in the soft X-ray bandpass, with a sensitive energy interval 0.2–12\,keV. 

Contemporaneous observations of 4U~1820$-$303 during both \ixpe observations were performed in the frame of the NICER GO Cycle 5 (proposal 6189 having PI Alessandro Di Marco). During the first observation, NICER observed the source for a total exposure of $\simeq$1.7\,ks (Observation ID 5050300117), while during the second observation, two data-sets were released (Observation ID 6689020101 and 6689020102) covering April 15 and 16 in several snapshots for a total exposure time of $\simeq$10.3\,ks and $\simeq$13.7\,ks respectively. 

The NICER data were processed with the NICER Data Analysis Software v010a released on 2022 December 16 provided under HEASOFT v6.31.1 with the CALDB version released on 2022 October 30. The background spectra have been estimated by applying the new SCORPEON model in the 0.22--15\,keV band. All the obtained spectra were grouped to have at least 30 counts per bin.

\subsection{Swift-XRT}

The Neil Gehrels Swift Observatory \citep{Gehrels2004} carries three instruments to enable the most detailed observations of gamma-ray bursts to date; the X-ray Telescope (XRT) is one of them, based on a sensitive, flexible, autonomous X-ray CCD imaging spectrometer. 
Swift-XRT coordinated observations with \ixpe for 4U~1820$-$303 were performed; they are used in this work to monitor the source status and to obtain spectral information during both the \ixpe observations. Given the source brightness, the Swift-XRT observations were performed in Windowed Timing mode (WT). Four Swift-XRT pointings covered the first \ixpe observation in the period 2022 October 11--13 (ObsIDs: 00014980028, 00014980029, 00014980030, 00014980031), while the second \ixpe observation consisted of 5 pointings on 2023 April 15 and 16  grouped in 3 data-sets (ObsIDs: 00014980055, 00014980056 and 00014980057). These data have been used to monitor the source HR along the two \ixpe observations, as reported in Figure~\ref{fig:hr}. The data were extracted using \texttt{xselect} v2.5b released with HEASOFT v6.31.1. The source and background extractions were performed using the imaging capabilities of Swift-XRT. Given the source counting rate $\ge$100\,count\,s$^{-1}$, data can be affected by pile-up as reported in the appendix of \citet{Romano2006}.

\subsection{NuSTAR}

\nustar, the Nuclear Spectroscopic Telescope Array, observatory \citep{Harrison2013} consists of two identical X-ray telescope modules, referred to as FPMA and FPMB, providing broadband X-ray imaging, spectroscopy, and timing in the energy range of 3–79\,keV with an angular resolution of 18\arcsec\ (FWHM) and spectral resolution of 400\,eV (FWHM) at 10\,keV
During both the \ixpe observations of 4U~1820$-$303, \nustar performed coordinated observations, one (ObsID: 90802327002) from 2022 October 12 at 14:31 UTC to October 13 at 2:41 UTC with a total exposure of $\simeq$16.9\,ks and other two on 2023 April 15 and 16 (ObsIDs: 90902308002 and 90902308004) with a total exposure of $\simeq$16.8\,ks and $\simeq$15.2\,ks, respectively.

The \nustar data were processed with the standard Data Analysis Software (\texttt{nustardas} 16Feb22 v2.1.2) provided under HEASOFT v6.31.1 with the CALDB version released on 2023 April 4. A circular 150\arcsec\ radius region has been used for both source and background spectra extractions, with the source region centered on the locations of 4U~1820$-$303 and the background one in an off-center sourceless region in the detector image. All the  obtained spectra were grouped to have at least 30 counts per bin.



%


\bibliography{biblio}{}
\bibliographystyle{aasjournal}



\end{document}